\documentclass[sigconf]{acmart}

\usepackage{booktabs}

\copyrightyear{2019}
\acmYear{2019} 
\setcopyright{iw3c2w3}
\acmConference[WWW '19 Companion]{Companion Proceedings of the 2019 World Wide Web Conference}{May 13--17, 2019}{San Francisco, CA, USA}
\acmBooktitle{Companion Proceedings of the 2019 World Wide Web Conference (WWW '19 Companion), May 13--17, 2019, San Francisco, CA, USA}
\acmPrice{}
\acmDOI{10.1145/3308560.3316485}
\acmISBN{978-1-4503-6675-5/19/05}

\usepackage{balance}

\usepackage{float}
\usepackage{graphicx}
\usepackage{amssymb,amsmath}
\usepackage[english]{babel}

\usepackage[T1]{fontenc}
\usepackage[utf8]{inputenc}

\usepackage{algorithm}
\usepackage{algpseudocode}

\usepackage{hyperref}

\usepackage{longtable}
\usepackage{graphicx,grffile}

\providecommand{\tightlist}{%
  \setlength{\itemsep}{0pt}\setlength{\parskip}{0pt}}

\begin{document}

\title{Detecting Areas of Potential High Prevalence of Chagas in Argentina}

\author{Antonio Vazquez Brust}
\affiliation{
   \institution{Fundación Bunge y Born}
}
\author{Tomás Olego}
\affiliation{
   \institution{Fundación Bunge y Born}
}
\author{Germán Rosati}
\affiliation{
   \institution{Fundación Bunge y Born, UNSAM}
}
\author{Carolina Lang}
\affiliation{
   \institution{Fundación Bunge y Born, UBA}
}
\author{Guillermo Bozzoli}
\affiliation{
   \institution{Fundación Bunge y Born}
}
\author{Diego Weinberg}
\affiliation{
   \institution{Fundación Mundo Sano}
}
\author{Roberto Chuit}
\affiliation{
   \institution{Fundación Mundo Sano}
}
\author{Martin A. Minnoni}
\affiliation{
   \institution{Grandata Labs}
}
\author{Carlos Sarraute}
\affiliation{
   \institution{Grandata Labs}
}

\renewcommand{\shortauthors}{A. Vazquez Brust et al.}


\begin{abstract}

A map of potential prevalence of Chagas disease (ChD) with high spatial disaggregation is presented. It aims to detect areas outside the Gran Chaco ecoregion (hyperendemic for the ChD), characterized by high affinity with ChD and high health vulnerability.

To quantify potential prevalence, we developed several indicators: an Affinity Index which quantifies the degree of linkage between endemic areas of ChD and the rest of the country. We also studied favorable habitability conditions for \emph{Triatoma infestans}, looking for areas where the predominant materials of floors, roofs and internal ceilings favor the presence of the disease vector.

We studied determinants of a more general nature that can be encompassed under the concept of Health Vulnerability Index. These determinants are associated with access to health providers and the socio-economic level of different segments of the population.

Finally we constructed a Chagas Potential Prevalence Index (ChPPI) which combines the affinity index, the health vulnerability index, and the population density. We show and discuss the maps obtained. These maps are intended to assist public health specialists, decision makers of public health policies and public officials in the development of cost-effective strategies to improve access to diagnosis and treatment of ChD.

\end{abstract}

\ccsdesc[300]{Applied computing~Health care information systems}
\ccsdesc[300]{Information systems~Data mining}

\keywords{Chagas disease; neglected tropical diseases; epidemics; health vulnerability; migrations; call detail records; social network analysis}

\maketitle


\section{Introduction}\label{introduction}

This document presents the processing criteria and the analysis techniques used to develop a  map with high spatial resolution, which allows us to identify areas of Argentina that are more likely to be inhabited by a population potentially affected by Chagas disease (ChD). In the absence of disaggregated information regarding the prevalence of the disease in each locality, and to inform public policies aimed at treating it, this instrument will allow a focused follow-up in the geographic areas that need it most.

The fundamental dimension of analysis is the \textit{potential prevalence of Chagas}, which will be operationalized through a \textit{Chagas Potential Prevalence Index} (ChPPI). It is defined as the \emph{proxy} indicator to determine whether a population lives in an area characterized by a high probability of being affected by \emph{Trypanosoma cruzi}.

The index combines a set of characteristics of the population, such as their mobile phone communications, and a series of variables and indicators linked to health coverage in the areas of residence. As will be explained later, there are grounds to consider an index of this kind as a promising alternative in countries that do not have complete epidemiological records.

The first source of data used were anonymized Call Detail Records (CDR),
which contain information about incoming and outgoing calls of users. This registry allowed the construction of an \textit{Affinity Index}, which determines the level of linkage of the resident population in the Argentine territory, particularly in endemic areas (that is, where a high prevalence of the ChD by proximity and contact with the natural vector, \textit{Triatoma infestans}, is observed) with the population living outside these areas.

On the other hand, the proximity of health centers and the socioeconomic level of the population as health determinants were used for the construction of a Health Vulnerability Index, which complemented the analysis of the Affinity Index. These determinants are closely related to the health status of a person, understood in a broad sense: biological, psychological and social health. If they are absent or of insufficient magnitude, a state of vulnerability is produced~\cite{chuit2}.

By combining both indices, we generated an indicator that seeks to identify areas characterized by high contact with areas where Chagas disease is endemic and where a high level of health vulnerability is registered.
Complementing the affinity information with health vulnerability information allows us to interpret the relationship between both and to understand holistically the phenomenon of infection and the local socioeconomic situation.

The information presented has a high level of spatial disaggregation. The minimum unit of analysis is the census block, the smallest statistical unit for which public socio-demographic information is available. 
The size of census blocks in urban areas is determined by the number of homes: the blocks cover an average of 300 homes. 
In Argentina, according to data from the 2010 Census, there are more than 52,000 census blocks distributed throughout the country.

The map, by pointing out the ``hot'' areas of high affinity with the endemic area and high health vulnerability, was conceptualized and developed as an input for different users: researchers, public health specialists, decision makers of public health policies and public officials, among other actors. In particular, for decision-makers, this map would facilitate the development of cost-effective strategies to improve access to diagnosis and treatment of ChD.

The document is structured as follows:
Section~\ref{related-work} summarizes some relevant background on the subject and marks the main advances regarding the works of
\cite{monasterio2016analyzing,monasterio2016uncovering,sarraute2015descubriendo}.
In  Section~\ref{affinity-chagas} the methodology used for the construction of the affinity index is presented, together with the processing of the information linked to the CDRs.
Section~\ref{health-vulnerability-definition} presents some fundamentals about the concept of health vulnerability and a detailed explanation of the different
procedures and techniques applied for the construction of the vulnerability index.
In Section~\ref{final-index}, the functional form and calculation of the \textit{ChPPI} is developed and justified, and in Section~\ref{results} the final map is presented and a first descriptive analysis is made from the results.
Finally, the limitations and future lines of research opened by this work are discussed in Section~\ref{conclusion}.


\section{Related Work}\label{related-work}

The continued emergence of large databases, added to the growing capacity of computer processing (the phenomenon called \emph{Big Data}), has produced great advances in different disciplines, including epidemiology. By leveraging billions of cellular call records, and under certain conditions, it is now possible to model the diffusion pattern of certain endemics~\cite{frias2011agent}.

Call record databases have emerged as a source of particular interest for the study of human migrations~\cite{candia2008uncovering}. A review of recent scientific literature shows an increasing body of work using call record analysis for the study of mobility patterns on different scales. The use of these new sources in mobility studies offers some notable advantages over traditional options such as origin-destination surveys.

Call records offer optimal geographical and population coverage~\cite{wang2018applying}, and with a much lower cost of acquisition than the significant resources required to carry out large-scale mobility surveys. As shown in \cite{kang2010analyzing}, the analysis of millions of call detail records allows them to infer patterns of mobility aggregated at different geographic scales, disaggregating certain characteristics by demographic group. On the other hand, the still novel nature of this source implies that its use as a research resource is in its initial and exploratory period. Given that the data is being collected for purposes other than scientific research, the presence of undocumented biases, noise and omissions is expected. In particular, biases depend on the cellular telephony coverage in the analyzed area and the market penetration of the different mobile phone companies.
In addition (in contrast to traditional mobility surveys) call records are extremely detailed in their spatial and temporal attributes, but much more superficial in the sociodemographic ones~\cite{wang2018applying}.

For this work, we had access to anonymized records of about 50 million calls per day between mobile phones, over several months. Distinct call patterns can be inferred, establishing their relationship with social phenomena such as seasonal or long-term migrations. Furthermore, if statistics were available at a detailed level (that is, at the highest possible level of disaggregation), the migratory patterns of individuals infected with parasites, viruses or bacteria could be inferred under certain circumstances.

In recent years, seminal analyses of mobile phone communications to detect potential risk areas for the Chagas disease were carried out in two Latin American countries (Argentina and Mexico). These works were presented in~\cite{sarraute2015descubriendo,monasterio2016analyzing,monasterio2016uncovering}.
These studies showed that geolocated call records are rich in social information and can be used to infer whether an individual has lived in an endemic area at some point in his life.

In this study, progress was made in two fundamental directions with respect to the aforementioned works:

(i) The affinity model between endemic and non-endemic areas previously used was further developed. Instead of assuming uniform probabilities of Chagas transmission in the endemic area (EA), complementary information was used about the homes located in EA in order to identify differentials in those probabilities of transmission. This point is addressed in Section~\ref{affinity-chagas}.

(ii) A more general dimension was incorporated which affects the probability of contagion of the disease: health vulnerability, developed in Section~\ref{health-vulnerability-definition}.

In the present work, the correlations found between communications, mobility, access to the health system, demographic characteristics and the distribution of Chagas disease are refined; and the project is scaled up at the national level.


\section{Affinity Index}\label{affinity-chagas}

\subsection{Motivation}

As previously mentioned, the Affinity Index quantifies the degree of linkage between endemic areas of Chagas disease in Argentina (also known as the emph{Gran Chaco Argentino}) and the rest of the country. This index integrates two differentiated dimensions that use different sources of data: telephone records, housing conditions, and location of health centers, among others.

The first dimension is linked to the affinity between endemic and non-endemic areas, that is, to what extent the different areas of the country are linked to endemic areas (EA). Given that outside the EA, there are no favorable conditions for the reproduction of the main vector of the disease (the species \textit{Triatoma infestans} or ``vinchuca''), one of the main vehicles for the spread of the disease is linked to migratory currents. This fact is even more relevant considering that many of the provinces that make up the Gran Chaco have been historical producers of labor for the Argentine economy, at least between the beginning and ends of the 20th century~\cite{busso}. Therefore, the possibility of detecting areas with a high potential prevalence of ChD requires the correct detection and mapping of migratory flows between the Gran Chaco and the rest of the country.

A first obstacle in this sense lies in the lack of demographic information with a level of disaggregation adequate for the objectives set. In effect, the classical source (and virtually the only source of the National Statistical System with national rural and urban coverage) for the study of internal migrations is the National Population Census. However, in general, they are limited to quantifying internal migration flows at the provincial level. Although the places of birth, habitual residence and previous residence (five years ago) are investigated, the data are published only at the provincial level.

These limitations entail the need to explore other sources of information to quantify migration flows. The quantification of the affinity between both types of zones is based on the assumption that a high connection of mobile communications (properly filtered and processed, as will be seen below) constitutes a valid approximate indicator of the existence of migration flows between both areas. A high degree of affinity between an endemic area and an external area could be considered as an indirect indicator of the presence of a migrant population from an endemic area that lives at the time of analysis in a non-endemic area.

In this way, quantifying the degree of affinity of each region with high spatial precision would allow identifying those areas where a population with a greater chance of having contracted Chagas resides, due to the dynamics of transmission in Argentina~\cite{chuit}. 
Thus, the infection would have occurred in a previous period, when said population lived in an endemic area, or transplacentally, as those areas with a population that comes from an endemic area will also be more likely to have infected women of childbearing age, who may be infected and transmit the disease to their progeny.

The second dimension constitutes an extension with respect to previous works \cite{monasterio2016analyzing, monasterio2016uncovering, sarraute2015descubriendo}, wherein the endemic area is considered as a homogeneous area.
However, the Gran Chaco is not an area of homogenous vector transmission because of its geographical condition, history of control actions and socio-demographic and living conditions of its inhabitants. Although there are no complete and reliable records, the transmission (or the annual incidence) is facilitated in part by poor infrastructure of homes and the surrounding areas~\cite{chuit3}.
For this reason, we sought to identify the affinity between the human habitat and the vector of the disease (vinchuca).

\subsection{Processing Mobile Phone Data}

The main input used for the construction of the Affinity Index is mobile phone activity information. The database used consists of geolocated antennas and a table of calls (\textit{Call Detail Records} or CDR), wherein each call is associated with the antenna used to make the connection.

The analyzed database has approximately 70,000,000 records per day, collected over a period of  5 consecutive months.
To account for interactions between users, these were identified using an encrypted number. 
In this way, the anonymity of users in the database is protected, while maintaining the ability to distinguish between different users.

The CDRs are organized in registers, and each one provides the following fields:
encrypted number of the originator, encrypted number of the destinatary, direction of the call (incoming or outgoing), date and time of the call, duration of the call in seconds, code of the cell tower used.

The CDR analysis and extraction of the affinity score with endemic Chagas areas, for each antenna, can be summarized in three sequential stages.

\textbf{Detection of \textit{home antenna} of each user.}
The first step consisted in the detection, for each user $ u $, of its \textit{home antenna} $ H_u $, with the aim of geolocating the information of the communications. Thus, $ H_u $ is defined as the antenna in which most calls of the corresponding user are recorded in the period studied, based on the set of calls made on weekdays at night. That is, from Monday to Thursday, between 8pm at night and 6am in the morning of the following day. This range was chosen assuming that it coincides with the time slot during which most of the people remain in their residence. If there is more than one antenna with a maximum number of calls, one of them was randomly selected.
The number of users for whom $ H_u $ was defined is around 15 million.

\textbf {Calculation of seed affinity for each antenna.}
In this step, for each antenna $ a $, the \textit{seed affinity} $ s_a $ was determined. The affinity depends on the characteristics of the houses in the area where the antenna is located (see Section~\ref{housing_conditions}). Therefore, it depends only on geographic and demographic attributes, and not those related to phone communication.

As input to the process, a partition was made by quartiles of an indicator of the housing material conditions, whose estimation is developed in Section~\ref{housing_conditions}. Each antenna contained within the ecoregion of the Gran Chaco corresponds to an affinity indicator equal to the quartile of housing conditions to which it belongs (an integer between 1 and 4 inclusive), while outside the Gran Chaco polygon the indicator is 0. In this way, the $ s_a $ of each antenna can take an integer value between 0 and 4.

\textbf{Calculation of affinity indicators for antennas using CDR.}
For each inhabitant $ u $ of any antenna in the country, the goal was to assign an affinity level based on their telephone communications in the social graph $ G $. 
In particular, the set $ V_u (G) $ of users that make up the \textit{neighborhood} of $ u $ in the social graph was calculated. Each neighboring node $ v \in V_u (G) $ modifies the $ u $ score according to the intensity of the edge $ (u, v) $, and the seed demographic indicators in the region of the antenna where $ v $ lives (i.e. 
$ s_{H_v} $).

Given the modified indicators for each user, we can add the results grouping by household antenna, to find a distribution of the affinity indicator for each of them.

First, for each user $ u $ his seed affinity was defined as $ s_u = s_{H_{u}} $
that is, each user was associated with the seed affinity of the area in which they live.

Then, given $ u $ and its neighborhood $ V_u (G) $, the affinity indicator $ s'_u $ of each user was defined as 
$ s'_u = {\max}_{v \in V_u (G)} \{ s_ {v} \} . $
This means that the affinity indicator modified by the social graph is obtained by calculating the seed affinity for each $ v \in V_u (G) $, and then taking the maximum of the affinities that were found. Note that this number is also an integer between 0 and 4.

In this way, a series of indicators was defined for each antenna $ a $, which is the information of how many users among the inhabitants of $ a $ have each of the values $ s'_u $.
We call $ H_{a, k} $ the subset of the inhabitants $ H_a $ of an antenna $ a $, such that its $ s' $ is $ k $, that is:
$$H_{a,k} = \cup_{(h \in H_a / s'_h = k)} \{h\} . $$

By reporting the number of users in $ H_ {a, k} $ for each of the values $ k \in {0, \ldots, 4} $, we obtain a summary of the distribution of affinity levels in each antenna $ a $ in particular. That is, for each antenna, the distribution of users is given by the tuple:
$$< a, | H_{a,0} |, | H_{a,1} |, | H_{a,2} |, | H_{a,3} |, | H_{a,4} |>$$

\subsection{Housing Conditions} 
\label{housing_conditions}

Given that it is possible to relax the assumption about the uniform transmission chances throughout the ecoregion of the Gran Chaco,
and considering the social component of the disease, the habitability conditions of the houses and their direct relationship with the conditions conducive to housing the vector, we propose to disaggregate the region into multiple sub-areas. Each of the subdivisions represents a quartile with respect to an indicator that measures the degree to which local housing materials favor the presence of \textit{Triatoma infestans}.

For this, we used information from the 2010 National Population and Housing Census~\cite{indec_censo1} and we constructed an index that quantifies for each house its viability conditions for lodging the \emph{vinchuca}.

The ranch houses whose construction materials include adobe walls without
plaster and roofs of cane, mud or straw favor the domiciliation of the vector~\cite{indec_calmat}.
We worked with the following variables:

\begin{itemize}
\tightlist
	\item Predominant material of the floors
	 \begin{itemize}
	 	\tightlist
\item Ceramic, tile, mosaic, marble, wood, carpeting
\item Cement or fixed brick
\item Soil or loose brick
\item Other
	 \end{itemize}
	 
\item Predominant material of the roof exterior
	 \begin{itemize}
	 	\tightlist
\item Asphalt cover or membrane
\item Tile or slab (without cover)
\item Slate or tile
\item Metal sheet (without cover)
\item Fiber cement or plastic sheet
\item Cardboard sheet
\item Reed, palm, board or straw with or without mud
\item Other
	 \end{itemize}
	 
\item Internal ceiling (yes / no)

\end{itemize}

The favorable habitability conditions for the \emph{vinchuca} are the following:
(i) Predominant material of floors is ``Soil or loose brick'';
(ii) Predominant material of roofs is ``Reed, palm, board or straw with or without mud'';
(iii) Absence of internal ceiling.

Since these are discrete and categorical measurement level variables, we used the  Multiple Correspondence Analysis as a dimensionality reduction technique to recover the latent variables; then the dimension with the greatest variability was obtained and the coordinates for that latent variable were recovered.

\subsection{Antenna Level Indices}

\begin{figure}[t]
\centering 
\includegraphics[width=0.45\linewidth]{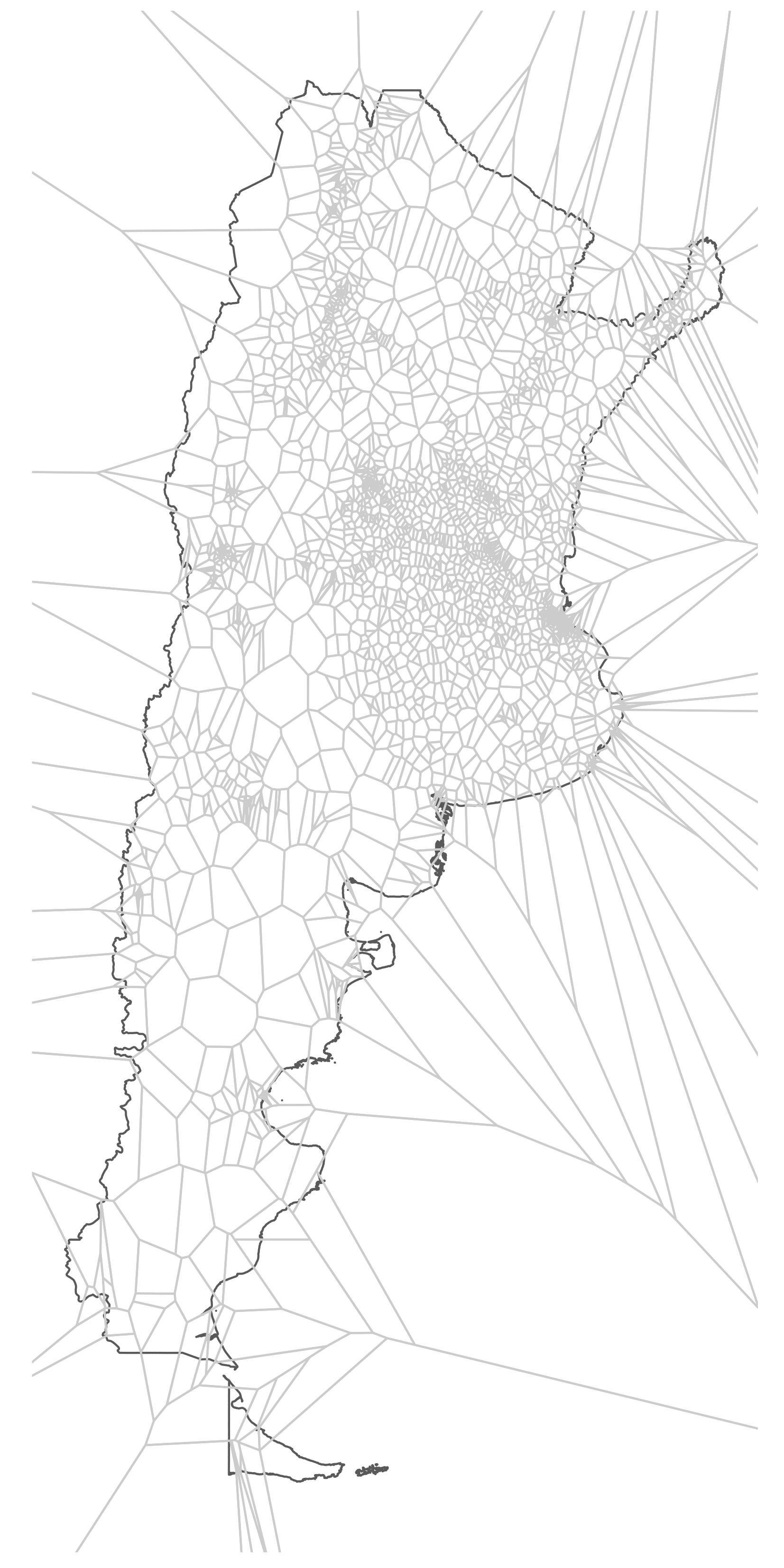}
\caption{Voronoi cells generated based on the position of the analyzed antennas}
\label{fig:voronoi}
\end{figure}

The information of the polygon of the endemic area has a level of resolution associated to the polygon of the census radius. Now, the information to generate the contact graph is associated with the antennas, that is, fixed points in space. Therefore, the first step for the generation of the antenna level indicators was to estimate the coverage area of the antennas in Argentina. For this purpose, a Voronoi diagram was used (shown in Fig.~\ref{fig:voronoi}).

The results of the housing quality index were added by each Voronoi cell, assigning a number of households based on the percentage of the area that intersected each census block. In other words, each block contributed a percentage of its values based on the proportion of surface shared with the Voronoi polygon or intersecting polygons.
Finally, the values obtained by antenna were graduated according to quantiles.


\section{Health Vulnerability Index}
\label{health-vulnerability-definition}

It is possible to state that the chances of contracting ChD, or of being infected by \textit{Trypanosoma cruzi}, are associated not only with the probabilities of contact with the vector or with movements or contacts with endemic areas. There are also determinants of a more general nature that could be encompassed under the concept of ``Health Vulnerability.'' These determinants are associated with access to services and health coverage of different segments of the population. By developing an index that quantifies this dimension, we can complement the Affinity Index with endemic areas of Chagas, since estimating the Health Vulnerability allows to prioritize ``hot zones'' outside the endemic area, those that have low access to sanitary services.

\subsection{Definition}

The existence of disparities in access to health services is a phenomenon that has been studied and documented on multiple occasions, highlighting evidence of disparities in access and constant medical coverage for different strata of the population. In general, the most impoverished population segments and/or residents in isolated areas have lower levels of access to these health benefits~\cite{beard, timyan1993access}.

In the literature on health problems, this notion of ``unequal access'' is often differentiated from the so-called ``vulnerability'' that refers, ultimately, to the risk of developing certain
diseases or the fact of being exposed to certain environmental risk factors. In this sense, the study of \emph{vulnerable populations} or \emph{vulnerability factors} is of interest.

An obstacle to the quantification of ``vulnerability''
emerges from the need to consider multiple factors that could explain inequalities in access to the health system.
The main indicators used by the studies analyzed in \cite{grabo}
are the condition of poverty, belonging to ethnic or racial minorities, the presence of chronic mental or physical illnesses and the lack of medical care. It should be noted that these indicators are defined at the individual level. There are also other factors that determine the level of health vulnerability that are more linked to the environmental dimension~\cite{pruss2016preventing}.

In our approach, several of the aforementioned indicators were considered. The proposed notion of health vulnerability is composed of the following associated factors:

(i) \emph{Access to health services and benefits from the state:} for this dimension, the proximity to health providers was used as the main indicator. A dataset was constructed that contains the location (latitude and longitude) of the vast majority of state health providers across the country. For this, sources from the National State and the Provincial States were integrated. The walking time was calculated from various points to the nearest Health Center.

(ii) \emph{Socio-economic Index of the population (SEI):} to construct the SEI, we used census information. Although this point is detailed below, it can be mentioned that the calculation of the indicator involved the processing of census information, corresponding to the CNPyV of year 2010. To this end, a series of relevant variables (educational level, indicators of unsatisfied basic needs, etc.) were selected and combined using \textit{variational autoencoders}, a method for dimensionality reduction based on neural networks (see Section~\ref{socioeconomic-index}).

\subsection{Methodology of Construction} 
\label{methodology-of-construction}

The dimensions mentioned above were combined to build a \emph{Health Vulnerability Map}. The objective of the vulnerability map is to identify areas with a potential deficit in the health coverage of the population, that is, that do not meet a minimum threshold in access to health services.
Taking into account this objective, a metric was constructed that allows ordering and classifying the different zones according to this potential deficit.

For the construction of the map, the following sources were collected and analyzed:
census data~\cite{indec_censo1}; 
census block polygons; 
location of public health providers: public hospitals, health centers and sanitary posts (in total, 16,564);
and the axes of streets (national and provincial routes, roads and urban traces) used to calculate by simulation distances between homes and health providers.

As mentioned, although data were processed at a lower disaggregation level (such as individual census data or data from health providers), for the construction of the final map, this information was added at the census block level .

The following sections detail the different procedures used for the preparation of the information to exploit population strata and the different processing and analysis techniques used.


\subsection{Accessibility to Hospitals and Health Centers}
\label{closeness-health-providers}

\paragraph{Construction and cleaning of the Health Providers dataset}

For the construction of the indicator \emph{Closeness to Health Providers}, the first step was the construction of a dataset with location records of the greatest possible number of health providers in the whole country, located with latitude and longitude coordinates.
This dataset was built from the integration of different sources of official data:

(i) National Base of Hospitals and Primary Care Centers: it was compiled by the Argentine Health Information System (\url{https://sisa.msal.gov.ar/}), and published by SEDRONAR on the IDERA site
(\url{http://catalogo.idera.gob.ar}). This dataset was used as the starting point and \emph{master base}. It was enriched and corrected based on information obtained from additional sources.

(ii) SUMAR program health providers. The information published on the site \url{http://programasumar.com.ar/efectores/} was collected via scraping.

(iii) List of hospitals and health care centers of the
National Program for Sexual Health and Responsible Procreation
(Ministry of Health). Data available by province. The data was downloaded and georeferenced. Source: \url{http://www.msal.gob.ar/saludsexual/centros.php}.

\paragraph{Classification of health providers by level
of complexity}

Subsequently, and as the last stage of the dataset cleaning process, the health providers were classified according to their level of complexity. We sought to reflect the fact that proximity to a highly complex hospital implies access to health benefits a priori greater than the proximity to a small health post. The type of problems, the emergencies attended and the attention provided by these establishments differ markedly. There were different classifications in the datasets used related to the notion of complexity, and the criteria varied among sources: the classifications were not homogeneous in the different lists of health providers consulted. 

A revision work was carried out with experts from the Mundo Sano Foundation, which allowed us to arrive at a classification that unifies the different denominations,
producing a simple classification into three categories, in decreasing order of complexity: (1) Hospital; (2) Health Center; (3) Sanitary Post.

After discarding the public health effectors that do not belong to any of the defined categories (for example, geriatric or administrative offices), 15,903 records of the 16,654 of the total collected were preserved.

\paragraph{Computing travel time to the nearest health center}

The next step was to compute the time necessary to reach the closest health care provider. From this point, the information was aggregated at the census block level.

We wanted to find the nearest public effector for each census block.
Since the shape, boundaries and surface
of the census blocks are very dissimilar throughout the country
(especially in rural or sparsely populated areas), we decided to calculate
the distances and times in the following way:

\begin{enumerate}
\def\labelenumi{\arabic{enumi}.}
\tightlist
\item
Within each block, 5 points (pairs of coordinates) were selected at random.
\item
The nearest health effector was identified for each point, using $k$-NN ($k$ nearest neighbors).
\item
The distance / time to the nearest health provider was calculated for each point.
\item
The 5 distances / times were averaged to obtain the final value.
\end{enumerate}

This procedure was performed for each category of health providers:  Hospital, Health Centers and Sanitary Posts.

For traveling time computations, we used Open Source Routing Machine (OSRM), a high-performance routing system that indicates the shortest route through public roads between any pair of source-destination coordinates~\cite{huber2016calculate}. 
To determine the routes, OSRM uses street grids downloaded from Open Street Map~\cite{OpenStreetMap}, a public repository of geographic information whose data quality has established it as a frequent source for mobility studies~\cite{haklay2010good,juran2018geospatial}.

\paragraph{Walking times}

The indicator used to measure access to health coverage was the walking time to the nearest health provider. This indicator is relevant given that there is evidence that, at least for certain types of medical treatments, walking distance to a health facility is a good predictor of the chances of completion of such treatment.
Indeed, the study \cite{beard} suggests that a distance greater than one
mile (approximately 1.6 km) results in a considerable increase in
the probability of not completing a rehabilitation treatment.
In turn, distances greater than 6.4 km result in a decrease in the average duration
of treatment in almost two weeks.

\begin{figure}[ht]
{\centering \includegraphics[width=0.98\linewidth]{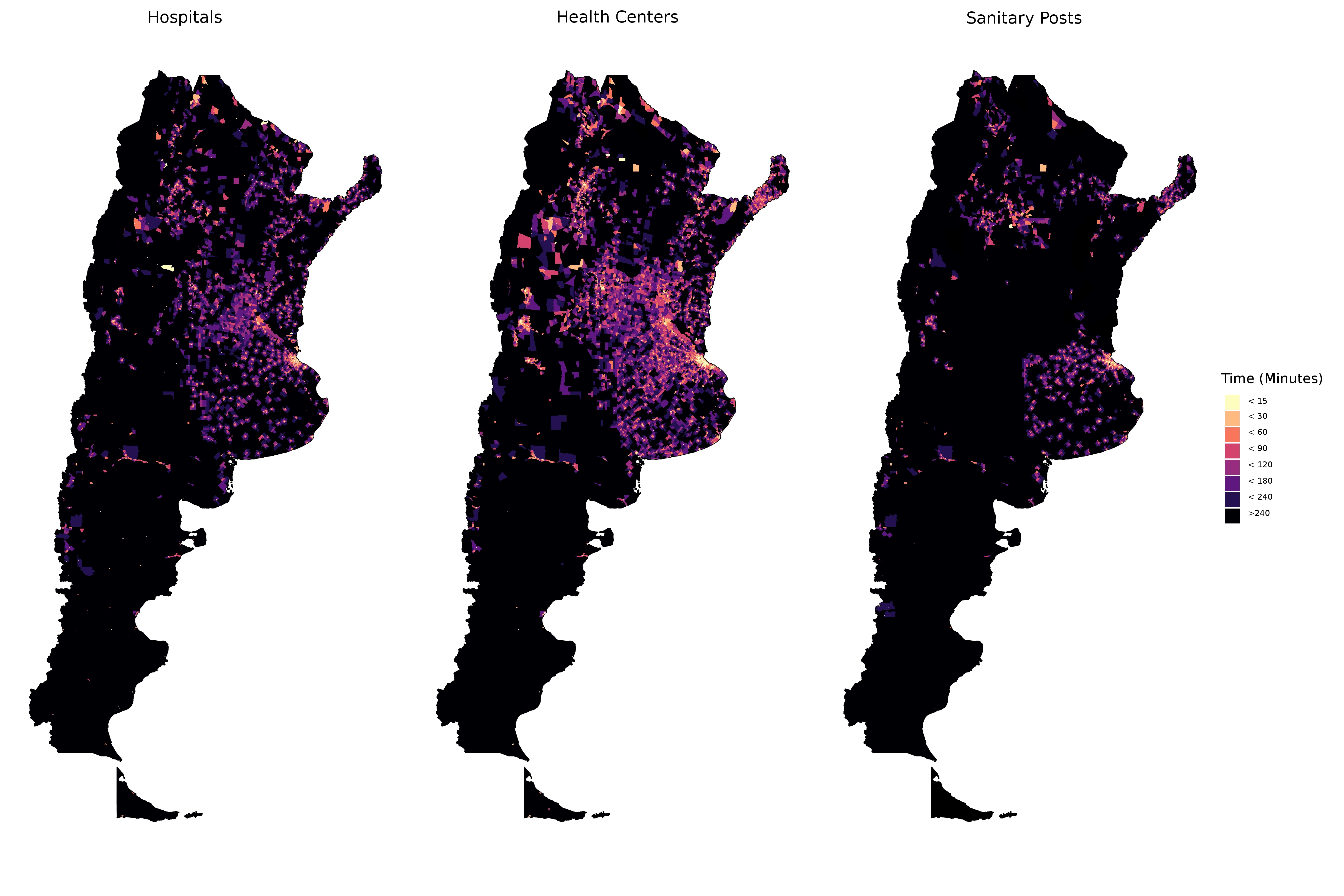}}
\caption{Walking times to health providers. Aggregated by census block.}
\label{fig:walking-time}
\end{figure}

Fig.~\ref{fig:walking-time} shows maps at the census block level of walking times to
Hospitals, Health Centers and Sanitary Posts for the whole country.
The final distance for each block  $ \Delta_{r} $ is the median distance between all the points sampled in each block, and includes all the distances to health providers in a census block $ r $.


\subsection{Socio-Economic Index}
\label{socioeconomic-index}

\paragraph{Input data}

To compute the \emph{Socio-Economic Index} (SEI), we used data from the 2010 Census. We worked with data at the individual level. Given that the SEI is usually a variable measured at the household level, we decided to calculate an index for each head of household in the Census dataset.

\begin{table}[ht]
\caption{Indicators used for the construction of SEI.}
\begin{tabular}{| l | l | }
\hline
Variable & Unit \\
\hline
Condition of home ownership & Housing \\
Quality of Materials & Housing \\
Quality of Connection to Basic Services & Housing \\
Quality of Construction & Housing \\
Overcrowding & Household \\
Unsatisfied Basic Needs (UBN) indicator & Household \\
Educational level of the Household & Household \\
Number of Unemployed in Household & Household \\
Existence of domestic services & Household \\
Activity condition & Individual (head) \\
Educational level & Individual (head) \\
\hline
\end{tabular}
\label{tab:sei-indicators}
\end{table}

To estimate the values of the index, we used the variables from Table~\ref{tab:sei-indicators}, which were ordinalized.
To build the index, a thermometer encoding was used for the ordinal variables. Let $ N $ be the number of cases and $ v_1, \ldots, v_I $ the variables. For each variable $ v_i $, there are \(K_i \) categories. The following coded variables were created $ x ^{(i)}_{k_i} $ for each variable $ v_i $ and for each category $ k_i $ where $ 2 \leq k_i \leq K_i $. In each case $ j $ with $ 1 \leq j \leq N $ it holds that:
\begin{equation}
x^{(i)}_{k_i}(j)=
    \begin{cases}
      0, & \ si \ v_i(j) < k_i \\
      1, & \ si \ v_i(j) \geq k_i
    \end{cases} 
\end{equation}

\paragraph{Construction of the final SEI}

For the construction of the final index, we used a dimensionality reduction technique called \textit{autoencoder}~\cite{goodfellow}. 
Autoencoders are an architecture based on
neural networks. In general, an autoencoder has the
objective of finding a representation of the input data
(\emph{encoding}), usually in order to perform a reduction
of dimensionality. Autoencoders work
by simply learning to replicate the inputs in the outputs. While
this seems a trivial problem, by introducing various restrictions to
the network, this task can become very complex. For example, you can
limit the size of the internal representation or add noise
to the inputs~\cite{geron}.

An autoencoder is composed of two parts:
an encoder (or \emph{recognition network}) that converts the inputs to
an internal representation, followed by
a decoder (or \emph{generative network}) that reconverts
the internal representation to the outputs.

\begin{figure}[th]
{\centering \includegraphics[width=0.90\linewidth]{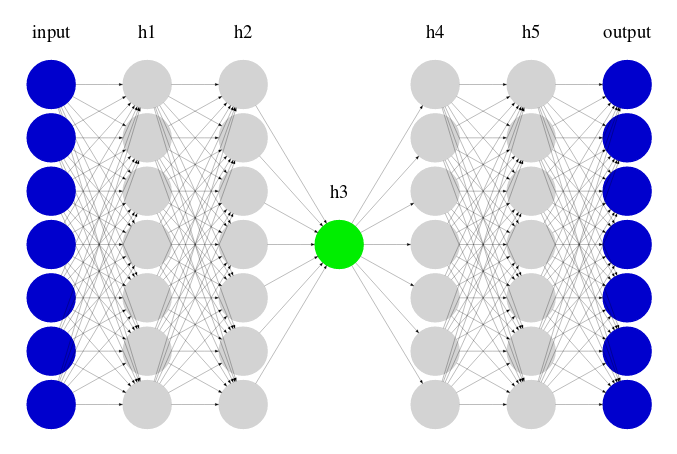} 
}
\caption{Scheme of the autoencoder used.}\label{fig:autoencoder}
\end{figure}

It usually has an architecture similar to a Multi-Layer Perceptron, with the
restriction that the number of neurons in the input and output layers must be
equal.
The trained model has as final layer a logistic function and
performs a \textit{dropout} of 0.5 in its intermediate layers to regularize and
achieve coefficients with good generalization capacity (see Fig.~\ref{fig:autoencoder}).

The model was trained with ADAM~\cite{geron} defining as data batches resampling with repetition on the empirical distribution of cases, to favor convergence.

As we worked with the whole population (since the data set is the census itself), the objective of the model was the generation of a descriptive measure. That is why the ability to explain the same population using the weighted average of the probability of each variable category was taken as a metric:
\begin{equation*}
 E(\hat{x}, x) = \frac{1}{N} \sum_{j = 1}^{N}  \sum_{i = 1}^{I} \sum_{k_i = 2}^{K_i} 
\frac{1}{\sum_\ell K_\ell}
e^{ x^{(i)}_{k_i}  \log \left(\hat{x}^{(i)}_{k_i} \right) + \left(1-x^{(i)}_{k_i} \right) \log \left(1-\hat{x}^{(i)}_{k_i} \right) } 
\end{equation*}
The final model retains 85 \% of the total input information.
In this way, starting at $ h_{3} $ each head of household, and therefore, each household, is classified with a value resulting from the autoencoder which we will call $ s_i $, the SEI.

From the SEI, an aggregate measure was generated for each block, based on the socioeconomic level of the heads of household. Thus, for the heads of household $ i $ with a socioeconomic level index $ s_i $ that live in the census block $ r $ with a population of $ n_r $ heads of household, we define a variable $ \eta $ as
\begin{equation}
    \eta_{r} = \frac{1}{4} Q_{.25}(\mathbf{s}_{r_{1,...,n_r}}) + \frac{1}{2} Q_{.5}(\mathbf{s}_{r_{1,..,n_r}}) + \frac{1}{4} Q_{.75}(\mathbf{s}_{r_{1,..,n_r}})
\end{equation}
where $ Q_{p} $ is the corresponding $ p $ quantile and $ \eta_ {r} $ is the result of applying the summary measure known as ``Tukey Trimean'' which achieves a compromise between robustness and efficiency compared to the median.

\begin{figure}[th]
{\centering \includegraphics[width=0.90\linewidth]{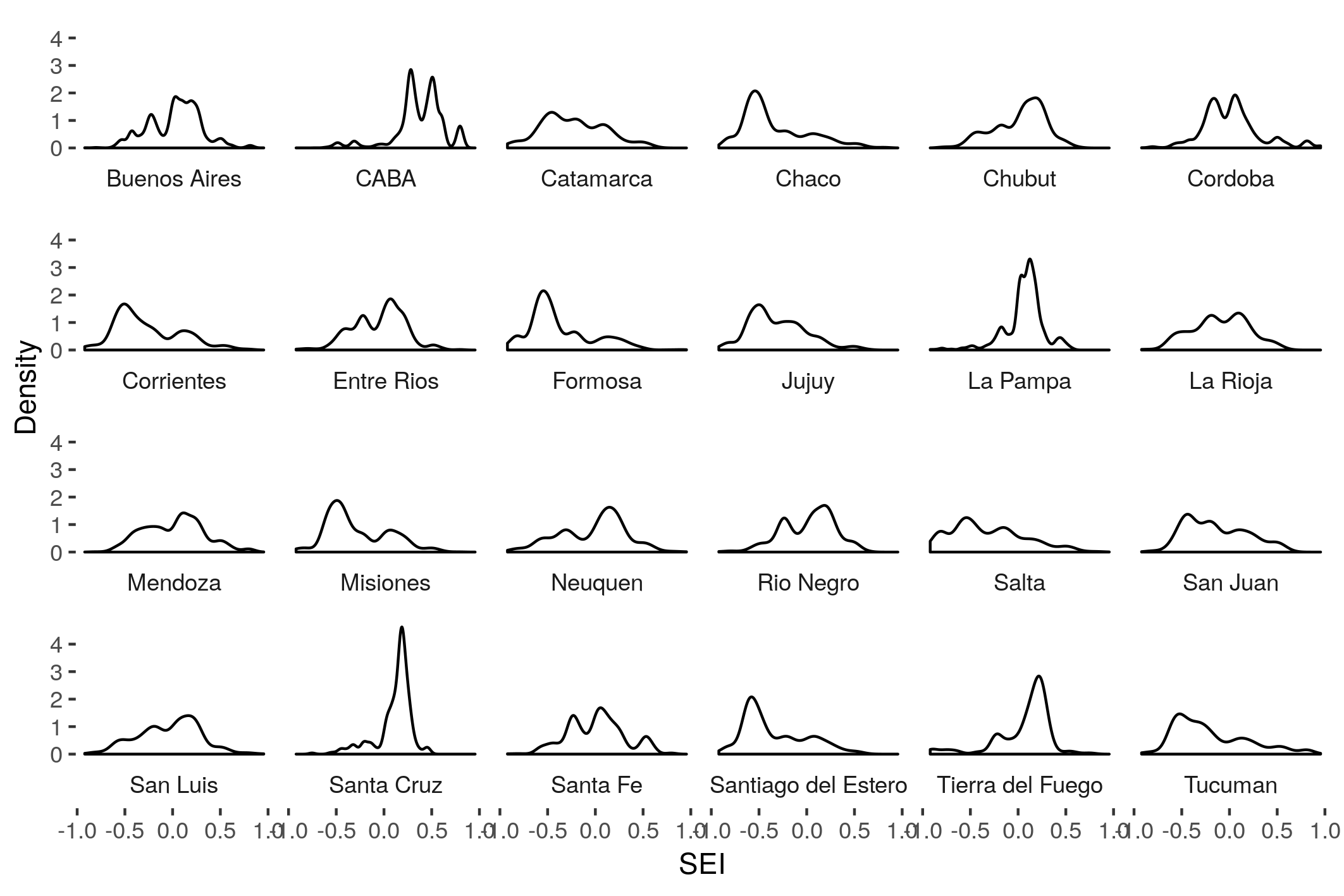} 
}
\caption{Density plot of the SEI by province.}
\label{fig:dens_plot_SEI_provs}
\end{figure}

As can be seen in Fig.~\ref{fig:dens_plot_SEI_provs}, the constructed index seems to capture to a large extent the provincial disparities:
the CABA presents a distribution clearly skewed to the right (higher values of the SEI);
on the other hand, provinces such as Chaco, Formosa, Jujuy and Salta have distributions skewed to the left.


\subsection{Generating the Health Vulnerability Index}
\label{generating-health-vulnerability-index}

To build the final Health Vulnerability Index, it was necessary to combine 
the socio-economic index and the accessibility to health centers  at the census block level. The strategy adopted to generate a composite index was that of principal components, that is, to look for the linear combination of the variables that can explain the maximum variance.

Now, when the distribution of the variables presents atypical distributions, that is multimodal, asymmetric and/or with heavy tails, the interpretation of the main components is difficult because the method is sensitive to the scale of the variables~\cite{jolliffe2006principal}. A possible solution is to transform the variables using the data ranks~\cite{BaxterRankPCA,compositionalRank}

Therefore, following \cite{solomon}, we sought to standardize each of the variables $ X_{j} $  by means of the rankit transformation:
\begin{equation}
    rankit(X_{i,j}) = \frac{r_j(X_{i,j}) - 0.5}{n}
\end{equation}
For each observation $ i $ of each variable $ X_{j} $ the rank $ r_j (X_{i, j}) $ (which varies between 1 and n) is calculated, then we subtract 0.5 and divide by the total of records $ n $.
In this way, the differences in units that could exist between variables were eliminated, and invariance was achieved respect to changes of scale, displacements and monotonous transformations.

Then, following \cite{EggerPCA,LiuPCA}, a semiparametric principal component analysis was performed (see Algorithm~\ref{SPCA}).

\begin{algorithm}
	\caption{Semiparametric Principal Components Analysis}\label{SPCA}
	\begin{algorithmic}[1]
		\Procedure{S-PCA}{X}
		\For{\texttt{j in variables}}
		\For{\texttt{i in cases}}
		\State $z_{i,j}=\Phi^{-1}\big(\frac{r_j(x_{i,j}) - 0.5}{n})\big)$ 
		\Comment{$\Phi^{-1}$ is the inverse of the gaussian f.d.a.}
		\EndFor
		\EndFor
		\State \textbf{compute} $\Sigma $ \Comment{Covariance matrix}
		\State \textbf{find} $U$, $S$ such that $\Sigma = \frac{1}{n} US^2U^t$ \Comment{SVD}
		\State \textbf{return} $ZU^t$, $S$ \Comment{Coordinates and eigenvalues}
		\EndProcedure
	\end{algorithmic}
\end{algorithm}

To perform the combination of both variables, the Spearman correlation was calculated on the rankits~\cite{LiuPCA} and the nonparametric correlation matrix was decomposed. We found that the main eigenvector absorbed 72\% of the variability of the rankits, and that it was oriented in the inverse direction of mutual growth. This first eigenvector was thus considered as the optimal combination between both variables. In this way, and through the successive application of nonparametric transformations, we generated an index that is pollution tolerant and, above all, independent of the measurement characteristics of each input variable.

As will be seen below, the interest lies in having an index between 0 and 1 whose distribution is homogeneous for all units of analysis. A transformation of the cumulative distribution function type estimated by logsplines~\cite{KOOPERBERG1991327} was again applied using AIC as a regularization criterion on the main direction, thus forming the index $ HV_ {r} $ for each block $r$.

\begin{figure}[th]
	{\centering \includegraphics[width=0.7\linewidth]{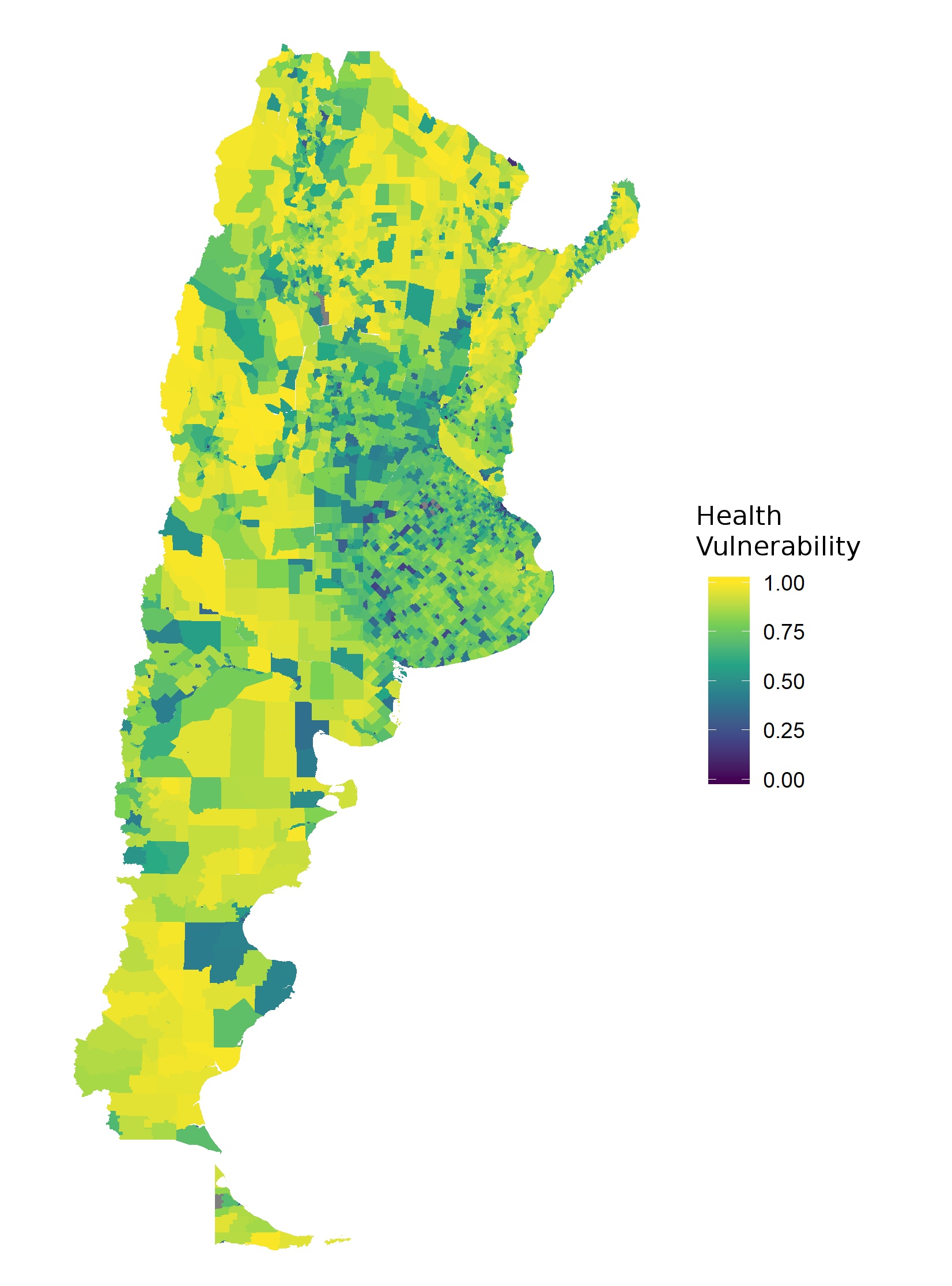} 
	}
	\caption{Health Vulnerability Index at the census fraction level for Argentina.}\label{fig:index_vs_fracc}
\end{figure}

Figure~\ref{fig:index_vs_fracc} shows a visualization of the Health Vulnerability Index for the whole country, aggregated by census fraction.
A census fraction is composed of a set of nearby census blocks~\cite{indec_censo2}.

We can detect two big zones: 
(i) a ``central'' region, essentially the Buenos Aires metropolitan area and the large urban agglomerations of each province, characterized by lower values of health vulnerability;
(ii) the rest of the country, mainly the less densely populated areas, with higher values.
However, even in the central region, there are zones with critical values.


\section{Generating the Combined Index}
\label{final-index}

\subsection{Population Density as Scaling Factor}

The ultimate goal of the Chagas Potential Prevalence Map is to guide actions and interventions by the State (at different levels). This implies that the concentration of the population in each of the census blocks is a scaling factor that needs to be considered in the final index.

For the design of cost-effective intervention strategies, it is necessary to quantify the potential impact in terms of the number of people affected. In this sense, it may be preferable to carry out actions in areas with a slightly lower potential prevalence index, but with a greater concentration of population in the territory: the cost per person treated may be lower in these areas of high population concentration, minimizing transportation costs.

That is why population density represents the final component of the index. In effect, for each block $ r $, density was calculated as the quotient between the number of inhabitants in the radius $ h_{r} $ and the total area in $ km ^ 2 $ of the census block $ a_{r} $. In order to combine it in the final index, the standardization strategy was again used by evaluating the cumulative probability distribution $ F $ on the variable, which removes both the unit differences and achieves invariance against changes in scale, displacements and monotonous transformations. The cumulative density function was estimated using logsplines:
\begin{equation}
	d_{r} = \hat{F}_d\left( \frac{h_{r}}{a_{r}} \right)
\end{equation}

\subsection{Chagas Potential Prevalence Index}

For the final construction of \textit{ChPPI}, the following variables were taken into account:
(i) local affinity with endemic area of Chagas,
(ii) health vulnerability, and 
(iii) population density.

From the point of view of public policies, with limited funds, it makes sense to prioritize areas with high population density and/or high ``affinity''. This composite index would allow, properly calibrated, to establish an order in which to expand health policies attending ChD, that take into account these different factors.

The final index was composed as follows:
\begin{equation}
	ChPPI_{r} = \frac
		{ {HV_{r}}^{\alpha} \ {d_{r}}^{\beta} \	AI_r }
		{ \frac{1}{R}\sum_{r=1}^R {HV_{r} }^{\alpha} \ {d_{r}}^{\beta} \ AI_r }
\end{equation}
where
${HV_{r}}^{\alpha}$ moderates the effect of the Health Vulnerability component with $ \alpha $ being the parameter that determines the impact;
${d_{r}}^{\beta}$  penalizes depopulated areas and $ \beta $ functions as the regulator;
and $AI_r$ is the Affinity Index for block $r$. In the denominator, $R$ is the total number of census blocks.

Note that $0 \leq \alpha,\beta$. When $\alpha,\beta \longrightarrow 0$, the effect of the Health Vulnerability or population density (respectively) are canceled.
On the other hand, when $\alpha,\beta \longrightarrow \infty$, 
the index is dominated by Health Vulnerability or population density (respectively).

Regarding $ ChPPI_{r} $, it should be interpreted as a relative index since its values indicate how much higher the potential prevalence of Chagas is in a census block compared to the median population value.


\section{Maps and Results}
\label{results}

The analysis of the results presented in this section intends to detect areas of interest for future field work and validation of the potential prevalence estimation model. That is, to try to find those areas in the country where the Chagas Potential Prevalence Index is high.
This indicates a high affinity --activity of cellular calls-- with endemic areas and high values in the health vulnerability index as an additional factor to prioritize a subsequent in-situ intervention. Both dimensions are combined in the final index.

For this analysis, only areas outside the Gran Chaco eco-region were considered --those where a high level of communication with the Gran Chaco is not attributable to the expected volume of calls among physically close persons.

However, we must take into account that the original map unit (census blocks) is not viable for the organization of field work: selecting relevant census blocks (with high affinity) and their corresponding contrasts would lead to a large territorial dispersion of the areas to visit. That is why the information was added at the local level.

Thus, after filtering blocks with less than 350 inhabitants, two metrics were built for each location:
(1) the average affinity along all the blocks that make up the locality (weighted by the population in each block);
(2) the average affinity of the blocks of the locality that were between the highest values of the national affinity distribution.

These metrics give rise to three types of possible situations:
(1) localities that have high affinity but ``dispersed'', exhibiting a similar concentration in all the blocks that compose it; we will call them \textit{high mean affinity};
(2) localities that have high and highly concentrated affinity values: that is, few blocks of the locality have high affinity values; named as locations of \textit{extreme blocks};
(3) locations that meet both criteria.

Such are the fundamental guidelines followed for the analysis. In addition, within each locality, we discarded the census blocks whose the population density does not reach 350 inhabitants per $ km^2 $, since it is necessary to reach a minimum local population concentration to guarantee the operative viability of a possible field intervention.

Within each province we selected localities of the above mentioned types.
\textbf{High affinity localities:} we selected the three with the highest average affinity weighted by the population of each block in each province.
\textbf{Localities of extreme blocks and localities with both types:} the same criterion was followed, the difference is that in this case, the average is calculated on the blocks that survive the previous filter; this may cause that some provinces do not present localities of these types.

\begin{figure}[ht]
	{\centering \includegraphics[width=\linewidth]{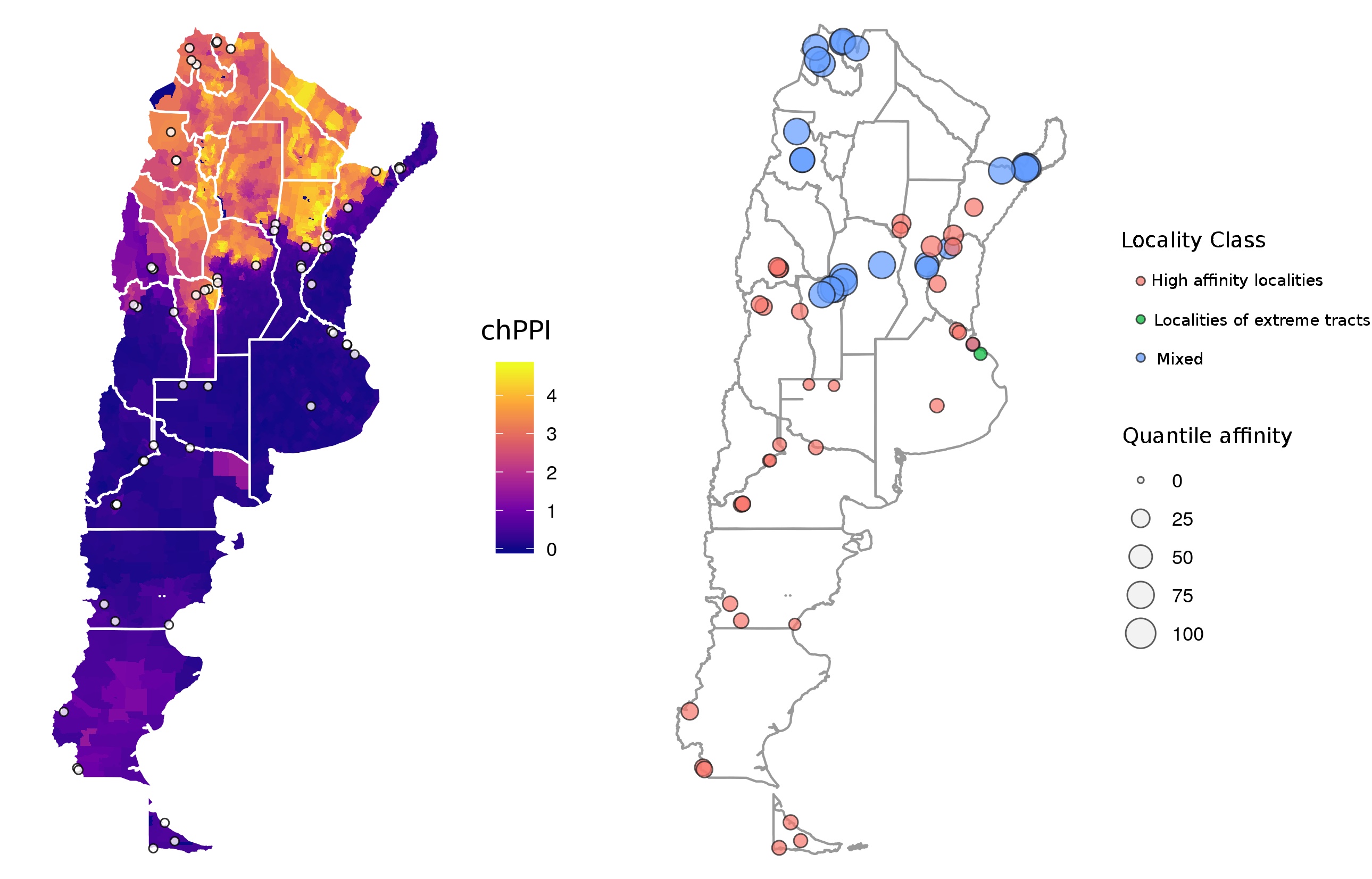}
	}
	\caption{Localities (homogeneous and concentrated) selected as potential areas of intervention according to levels of Potential Prevalence Index (ChPPI) and according to affinity percentile.}
	\label{fig:mapa_localidades_total}
\end{figure}

Figure~\ref{fig:mapa_localidades_total} shows the selected localities, projected on the map of the Potential Prevalence Index. As established in \cite{monasterio2016uncovering}, the areas with the highest affinity indexes are concentrated in the Metropolitan Region of Buenos Aires (including the city of La Plata) and in the provinces of Patagonia -- Neuquén, Río Negro, Chubut, Santa Cruz and Tierra del Fuego. Some localities of border areas to the Gran Chaco also appear as relevant, in provinces such as Misiones, Entre Ríos and Mendoza.

A first point to note is that there is only one location (La Plata, Buenos Aires) that is characterized by presenting extreme blocks but not high average affinity. At the same time, it can be noted that between the border areas of the Gran Chaco, those that combine extreme blocks and localities of high average affinity predominate within the selected localities. These localities are characterized (as expected) by higher values in affinity percentiles.

On the other hand, in the central and Patagonia zones, the situation seems to be different: the selected localities are characterized by being of high mean affinity and by lower values in affinity percentiles.

Although they share high affinity levels, the zones do not exhibit similar patterns in terms of their health vulnerability. In this dimension, the variability is high, covering areas in the entire range from low vulnerability in relation to the rest of the country (CABA, Vicente Lopez, La Plata), up to high levels: El Durazno and Jacipunco (Catamarca), Pueblo Libertador (Corrientes), Coranzulis (Jujuy), Poscaya (Salta).

\begin{figure}[ht]
{\centering \includegraphics[width = 0.90\linewidth]{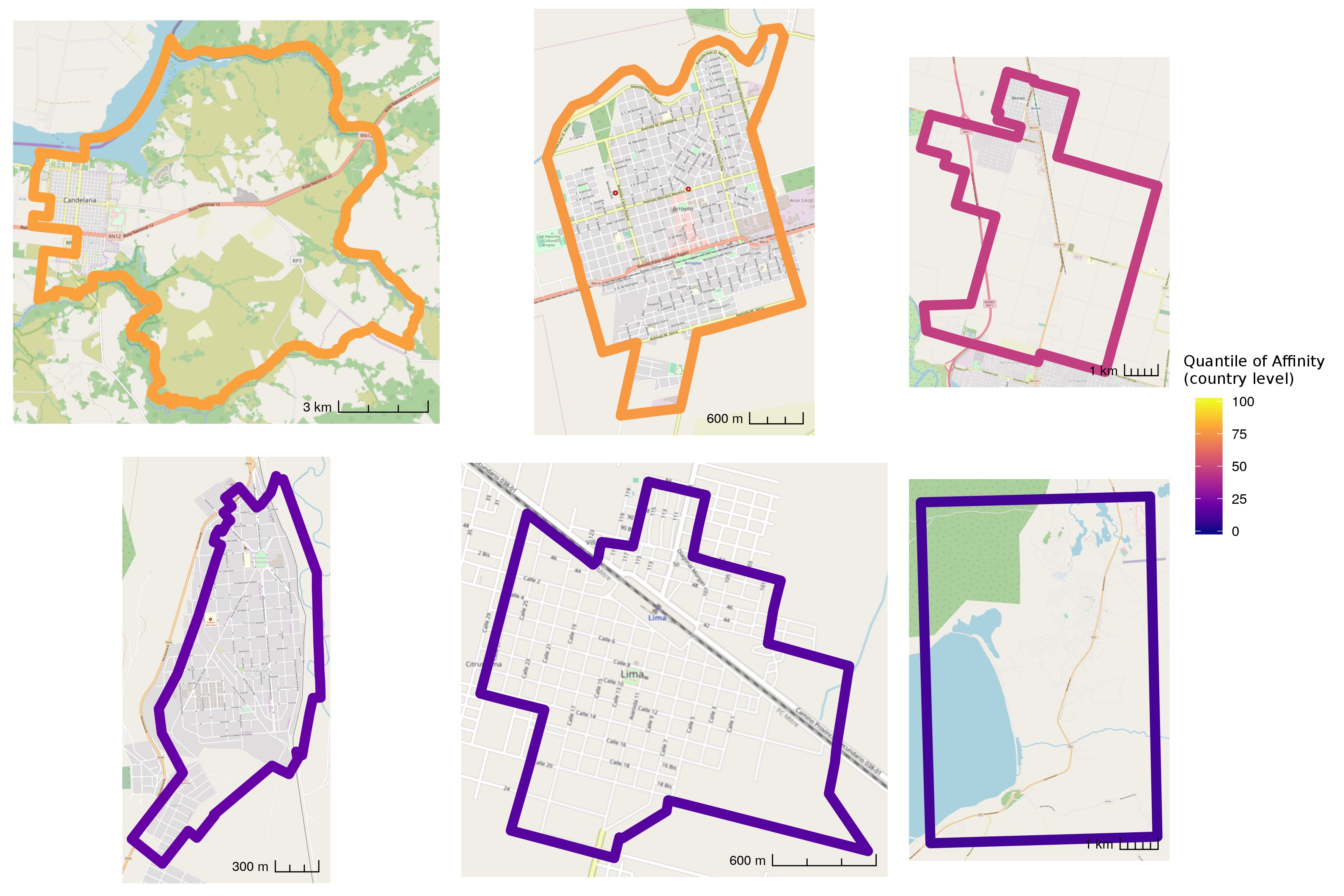}
}
\caption{From left to right, and from top to bottom: 
Candelaria (Misiones), Arroyito (Córdoba), Recreo (Santa Fe), 28 de Noviembre (Santa Cruz), Lima (Buenos Aires), Tolhuin (Tierra del Fuego).}\label{fig:mapa_localidades_zoom}
\end{figure}

Among the selected localities, in Fig.~\ref{fig:mapa_localidades_zoom} we zoom in on 6 localities with a population greater than 5,000 habitants, each in a different province and in decreasing order of affinity. As expected, due to its geographical proximity to the Endemic Area (EA), the localities in Misiones, Córdoba and Santa Fe show a greater affinity compared to their counterparts in the Center and South of the country. However, we observe that the affinity does not decrease homogeneously, i.e. in a continuous gradient as the localities move away from the endemic area. On the contrary, localities were detected in the Province of Buenos Aires and in Patagonia whose degree of affinity is much higher than population centers in provinces closer to the EA such as La Pampa. This suggests the existence of considerable migrations from endemic regions to the highlighted localities.


\section{Discussion and Future Work}
\label{conclusion}

A first result of the generation of this map involves opening some lines of work that may be of interest. The most obvious of them involves a process of revision, improvement and update of the data sources. 

On the one hand, the sources of information used could be expanded by incorporating records obtained over several years to identify dynamics related to migration patterns. 
In this work we used the volume of calls from residents of non-endemic areas with the endemic area as a proxy indicator of the existence of movements and population contacts between both types of zones. This implies an inferential leap: it is assumed that communication is a good indicator of the existence of migratory flows from endemic areas. In turn, a second assumption is that part of the migrant population is or was infected, at least, with a probability comparable to that of the non-migrant population.

Although this is a reasonable assumption, it is also expected that the same CDRs will allow --based on new processing-- to estimate effective migratory patterns and volumes. That is, from the succession of user connections to different antennas, real population movements could be mapped between endemic and non-endemic areas.

On the other hand, the CDR data have limitations that need to be pointed out. Perhaps the most important one is linked to the degree of coverage of the mobile phone antennas. In effect, this variable alters the spatial granularity of the analysis. In areas of low density, the greater distance between antennas makes the area assigned to each one more extensive, and increases the difficulty of identifying the underlying census block. 
Therefore, it is recommended as future work to perform a spatial analysis of conglomerates in order to identify homogeneous areas of blocks to mitigate said problem.

Regarding the use of additional sources integrated in the Map (especially the Health Vulnerability), a second iteration in the cleaning and consolidation of the georeferenced data of health providers emerges as a necessary task. This task would imply the incorporation of sources that were not used in this first approach. In turn, the classification according to complexity level of health care providers deserves a review.

The incorporation of new sources of data and dimensions linked to environmental risk and vulnerability are also of potential interest for future work.

Another valuable result of this work is to generate a replicable work methodology, 
that can be applied in other areas. The quantification of interactions between dispersed geographic areas, as well as the indicators of Health Vulnerability, can be considered as transversal dimensions that affect the evolution and the transmission of other pathologies besides the ChD. Therefore, a second line of future work is the possibility of considering maps such as the one described here (and subsequent updates) as an input for the study of other infectious diseases.


\bibliographystyle{ACM-Reference-Format}
\balance 

\bibliography{biblio}

\end{document}